\documentclass[preprint,12pt]{elsarticle}

\usepackage{amssymb}

\usepackage{lineno}

\usepackage{xspace}	
\usepackage{color}

\newcommand{\sqrts}[1]{\mbox{$\sqrt{s_{_{NN}}}$} = #1 GeV\xspace} 
\newcommand{\pt}{\mbox{$p_{\rm T}$}\xspace} 
\newcommand{\gevc}{\mbox{GeV/$c$}\xspace} 
\newcommand{\npart}{\mbox{$N_{\rm part}$}\xspace} 
\newcommand{\ncoll}{\mbox{$N_{\rm coll}$}\xspace} 
\newcommand{\pp}{\mbox{$p$~+~$p$}\xspace} 
\newcommand{\AuAu}{\mbox{Au~+~Au}\xspace} 
\newcommand{\UU}{\mbox{U~+~U}\xspace} 
\newcommand{\dNchdy}{\mbox{$dN_{\rm ch}/dy$}\xspace} 
\newcommand{\dNchdeta}{\mbox{$dN_{\rm ch}/d\eta$}\xspace} 

\journal{Physics Letter B}

\begin{document}

\begin{frontmatter}



\title{
  Predictions of Elliptic flow and nuclear modification factor 
  from 200 GeV \UU collisions at RHIC
}


\author[lbl]{Hiroshi Masui\corref{cor1}}
\ead{HMasui@lbl.gov}

\author[vecc]{Bedangadas Mohanty}
\ead{bedanga@rcf.bnl.gov}

\author[lbl]{Nu Xu}
\ead{NXu@lbl.gov}

\cortext[cor1]{Corresponding author}

\address[lbl]{Lawrence Berkeley National Laboratory, Berkeley, California 94720, USA}
\address[vecc]{Variable Energy Cyclotron Centre, Kolkata 700064, India}

\begin{abstract}
  Predictions of elliptic flow ($v_2$) and nuclear modification factor ($R_{AA}$)
  are provided as a function of centrality in \UU collisions at \sqrts{200}.
  Since the $^{238}$U nucleus is naturally deformed,
  one could adjust the properties of the fireball, density and duration of the hot and dense system, for example, 
  in high energy nuclear collisions by carefully selecting the colliding geometry.
  Within our Monte Carlo Glauber based approach, 
  the $v_2$ with respect to the reaction plane $v_2^{\rm RP}$ in 
  \UU collisions is consistent with that in \AuAu collisions, while the $v_2$ 
  with respect to the participant plane $v_2^{\rm PP}$
  increases $\sim$30-60\% at top 10\% centrality which is attributed to the 
  larger participant eccentricity at most central \UU collisions.
  The suppression of $R_{AA}$ increases and reaches $\sim$0.1 at most 
  central \UU collisions that is by a factor of 2 more suppression
  compared to the central \AuAu collisions
  due to large size and deformation of Uranium nucleus.

\end{abstract}


\end{frontmatter}


\section{Introduction}
\label{sec:introduction}

  Most striking findings at RHIC are the large elliptic flow $v_2$~\cite{Ackermann:2000tr}
  and the strong suppression of nuclear modification factor $R_{AA}$~\cite{Adcox:2001jp}.
  The $v_2$ is defined by the second harmonic Fourier coefficient of 
  azimuthal particle distribution with respect to the reaction plane,
  and the $R_{AA}$ is defined by the ratio of invariant yield in A + A collisions
  to that in p + p collisions scaled by number of collisions.
  Recent systematic measurements of $v_2$~\cite{Abelev:2008ed}
  as well as developments of viscous hydrodynamical models~\cite{Dusling:2007gi,Song:2007ux,Luzum:2008cw,Huovinen:2008te}
  provide a conservative upper limit of the viscosity $\eta$ to the entropy $s$ ratio $\eta/s \le 0.5$.
  This corresponds to the 6 times larger value of an absolute lower bound $\eta/s = 1/4\pi$
  predicted by strongly coupled gauge filed theories based on the AdS/CFT 
  correspondence~\cite{Policastro:2001yc,Kovtun:2004de}.
  It has been observed that the ratio $v_2/\varepsilon$ in different systems 
  from AGS to RHIC scale like $1/S \dNchdy$~\cite{Voloshin:2007af}
  as it was predicted by a low density limit of $v_2$~\cite{Heiselberg:1998es,Voloshin:1999gs},
  where $\varepsilon$ is the initial geometrical anisotropy (eccentricity),
  $S$ is the transverse area and \dNchdy is the charged particle rapidity density.
  The saturation of $v_2/\varepsilon$ would indicate that 
  the system is approaching the hydrodynamical limit and the collectivity no longer 
  increases when the system size becomes larger. 
  The measurements of transverse momentum spectra of charged hadrons showed that 
  the yield at most central \AuAu collisions at \sqrts{200} is suppressed by a factor 
  of $\sim$5 compared to the \pp reference scaled by number of binary 
  collisions $N_{\rm coll}$~\cite{Abelev:2006jr, Abelev:2007ra}
  and the similar level of suppression persists for neutral pions up to \pt = 20 \gevc~\cite{Adare:2008qa}.
  The integrated $R_{AA}$ above \pt $>$ 5 \gevc and $>$ 10 \gevc decreased 
  monotonically as a function \npart and there were no sign of saturation~\cite{Adare:2008qa}.

  Assuming the underlying dynamics remains the same, we ask what would happen 
  to $v_2$ and $R_{AA}$ for a larger colliding system $^{238}$U~+~$^{238}$U collisions ?
  Comparing to the $^{197}$Au nucleus, the $^{238}$U has a much larger mass and, 
  more importantly, it is largely deformed. 
  The planned \UU collisions at RHIC will be important for us to understand how
  those observables behave at higher particle density.
  Monte Carlo Glauber simulations showed that 
  the transverse number density $1/S \dNchdy$
  increases $\sim$ 35\% at most central events in ideal tip-tip collisions 
  (head-on collisions along the longest axes)~\cite{Nepali:2006ep}.
  The \UU collisions will become possible when the Beam Ion Source 
  becoming operational in 2012~\cite{Alessi:2005ve}.

  In this letter, we will report a geometrical approach based on the Monte Carlo Glauber 
  model to predict the elliptic flow $v_2$ as well as the nuclear modification factor $R_{AA}$ 
  in \UU collisions at top RHIC energy.
  In the Section~\ref{sec:glauber}, we will discuss our parameterization of Glauber model
  and define geometrical quantities which are used in this study.
  In the Section~\ref{sec:results}, the results of $v_2$ and $R_{AA}$ in \UU collisions will be presented 
  and compared to the data in \AuAu collisions at \sqrts{200}.

\section{Glauber Model}
\label{sec:glauber}

  The nucleon density distribution is parameterized by a deformed Woods-Saxon profile~\cite{Hagino:2006fj}
  \begin{eqnarray}
    \rho & = & \frac{\rho_0}{1 + \exp{([r - R']/a)}},  \label{eq:deformed_ws} \\
      R' & = & R \left[1 + \beta_2Y_2^0(\theta) + \beta_4Y_4^0(\theta)\right],
  \end{eqnarray}
  where $\rho_0$ is the normal nuclear density,
  $R$ and $a$ denote the radius of nucleus and the surface diffuseness parameter, respectively.
  We have used $R = 6.38$~fm and $a = 0.535$~fm for $^{197}$Au nucleus, 
  and $R = 6.81$~fm and $a = 0.55$~fm for $^{238}$U nucleus. 
  The $Y_l^m(\theta)$ denotes the spherical harmonics and $\theta$ is the polar angle 
  with the symmetry axis of the nucleus.
  Deformation parameters are $\beta_2 = 0.28$~\cite{Heinz:2004ir} and $\beta_4 = 0.093$~\cite{Moller:1993ed} for Uranium.
  The presence of $\beta_4$ modifies the shape of Uranium compared to 
  that only with $\beta_2$, which was implemented in several different models~\cite{Nepali:2006ep, Heinz:2004ir}.
  The radius increases $\sim$6\% (3\%) at $\theta = 0$ ($\theta = \pi/2$),
  while it decreases $\sim$3\% around $\theta = \pi/4$.
  We have assumed that Au nucleus is spherical ($\beta_2 = \beta_4 = 0$), thus 
  Eq.~(\ref{eq:deformed_ws}) reduces the spherical Woods-Saxon profile. 
  Recent calculation~\cite{Peter:2009}
  shows that the ground-state deformation of $^{197}$Au affects the eccentricity 
  of initial geometry overlap only at most central collisions 
  from both optical and Monte Carlo Glauber simulations.
  The positions of nucleons are sampled by $4\pi r^2 \sin{(\theta)} \rho(r) ~ d\theta d\phi$,
  where the absolute normalization of $\rho(r)$ is irrelevant.

  Both projectile and target U nuclei are randomly rotated along
  the polar and azimuthal directions event-by-event with the probability 
  distribution $\sin{\Theta}$ and uniform distribution for $\Theta$ and $\Phi$, respectively.
  The $\sin{\Theta}$ weight needs to be implemented to simulate unpolarized 
  nucleus-nucleus collisions.
  The results are averaged over all possible orientations unless otherwise specified.

  A binary nucleon-nucleon collision take places if 
  \begin{equation}
    d \le \sqrt{\frac{\sigma_{NN}}{\pi}},
  \end{equation}
  where $d$ is the distance between nucleons in the transverse direction orthogonal to the beam axis, 
  and $\sigma_{NN} = 42$ mb is the inelastic nucleon-nucleon cross section at $\sqrt{s}$ = 200 GeV.
  For each event, the total number of binary collisions \ncoll is calculated by the sum of 
  individual number of collision and the total number of participant nucleons \npart is 
  the number of nucleons that interacts at least once.

  Charged particle pseudorapidity density is obtained by a two component model~\cite{Kharzeev:2000ph}
  \begin{equation}
    \frac{dN_{ch}}{d\eta} = n_{pp} \left[ (1-x) \frac{\npart}{2} + x \ncoll \right],
  \end{equation}
  where $n_{pp} = 2.29$ and $x = 0.145$ are fixed to reproduce the PHOBOS results~\cite{Back:2004dy}.
  Event-by-event multiplicity fluctuations have been taken into account 
  by convoluting Negative Binomial Distribution for a given \npart and \ncoll
  \begin{equation}
    P(\mu, k; n) = \frac{\Gamma{(n+k)}}{\Gamma{(n+1)}\Gamma{(k)}} \left(\frac{\mu}{k}\right)^n
    \left( 1 + \frac{\mu}{k} \right)^{-(n+k)},
  \end{equation}
  where $\mu = n_{pp}$ is the mean of the distribution and $1/k = 0.5$ corresponds to deviation 
  from a Poisson distribution. 
  In this study, we have generated 1 million events for \UU collisions
  by randomly selecting an impact parameter $b$ according to the $d\sigma/db = 2\pi b$.

  \begin{figure}[htbp]
    \includegraphics[width=\linewidth]{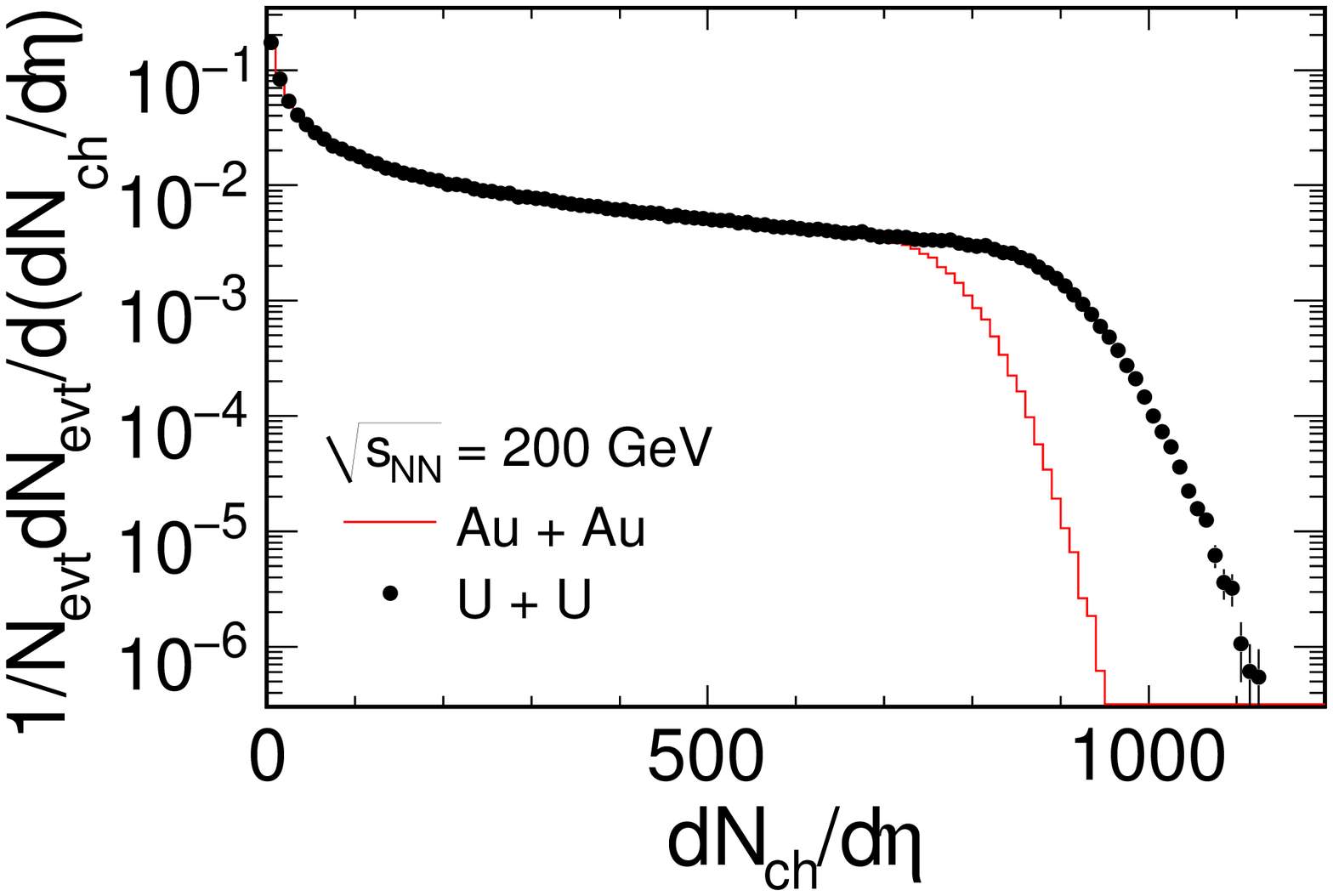}
    \caption{\label{fig:fig1}
      (Color online) \dNchdeta distribution in \AuAu (solid line) and \UU collisions (filled circles) 
      at \sqrts{200} by averaging over all orientations.
    }
  \end{figure}
    Figure~\ref{fig:fig1} shows the comparison of \dNchdeta distributions 
  in \AuAu and \UU collisions at \sqrts{200} from our Monte Carlo Glauber model. 
  The maximum \dNchdeta in \UU collisions increases $\sim$15\% compared to that in \AuAu collisions.
  We have defined the event centrality bins by the fraction of events in \dNchdeta. 
  The centrality bins are summarized in Table~\ref{tab:summary_centrality}.

  \begin{table*}[htbp]
    \begin{center}
    \caption{\label{tab:summary_centrality}
      Summary of centrality bins based on the \dNchdeta 
      and $\langle\npart\rangle$, $\langle\ncoll\rangle$, $\langle S_{\rm RP}\rangle$, $\langle S_{\rm PP}\rangle$,
      $\langle\varepsilon_{\rm RP}\rangle$, $\varepsilon_{\rm PP}\{2\} \equiv \sqrt{\langle\varepsilon_{\rm PP}^2\rangle}$ 
      and $\langle L\rangle$ for each centrality bin in \UU collisions at \sqrts{200}.
      Errors denote systematic uncertainties, see texts for more details of systematic error evaluations.
    }
    {\small
    \begin{tabular}{p{3.5em}p{2.5em}ccccccc}
    \hline
      centrality &  \dNchdeta  & $\langle\npart\rangle$ & $\langle\ncoll\rangle$ & $\langle S_{\rm RP}\rangle$ & $\langle S_{\rm PP}\rangle$ & $\langle\varepsilon_{\rm RP}\rangle$ & $\varepsilon_{\rm PP}\{2\}$ & $\langle L\rangle$ (fm) \\
    \hline
      0-5\%    &  $\geq$ 740  & 418 $\pm$ 6  & 1341 $\pm$ 105 & 30.9 $\pm$ 1.7 & 29.7 $\pm$ 1.7 & 0.021 $\pm$ 0.007 & 0.156 $\pm$ 0.004 & 4.4 $\pm$ 0.1\\
      5-10\%   &  $\geq$ 609  & 358 $\pm$ 14 & 1058 $\pm$ 52  & 27.1 $\pm$ 1.9 & 26.9 $\pm$ 1.9 & 0.08  $\pm$ 0.02  & 0.18  $\pm$ 0.01  & 4.2 $\pm$ 0.2\\
     10-20\%   &  $\geq$ 410  & 281 $\pm$ 13 & 751  $\pm$ 49  & 22.9 $\pm$ 1.8 & 22.7 $\pm$ 1.8 & 0.15  $\pm$ 0.02  & 0.24  $\pm$ 0.02  & 3.9 $\pm$ 0.1\\
     20-30\%   &  $\geq$ 269  & 199 $\pm$ 14 & 462  $\pm$ 45  & 18.4 $\pm$ 1.6 & 18.2 $\pm$ 1.6 & 0.23  $\pm$ 0.03  & 0.31  $\pm$ 0.03  & 3.5 $\pm$ 0.1\\
     30-40\%   &  $\geq$ 170  & 137 $\pm$ 14 & 272  $\pm$ 39  & 14.8 $\pm$ 1.5 & 14.5 $\pm$ 1.6 & 0.29  $\pm$ 0.03  & 0.38  $\pm$ 0.03  & 3.2 $\pm$ 0.2\\
     40-50\%   &  $\geq$ 101  & 89  $\pm$ 13 & 149  $\pm$ 31  & 11.8 $\pm$ 1.5 & 11.4 $\pm$ 1.5 & 0.34  $\pm$ 0.04  & 0.45  $\pm$ 0.04  & 2.9 $\pm$ 0.2\\
     50-60\%   &  $\geq$  56  & 55  $\pm$ 11 & 75   $\pm$ 22  & 9.3  $\pm$ 1.5 & 8.8  $\pm$ 1.5 & 0.38  $\pm$ 0.04  & 0.51  $\pm$ 0.05  & 2.6 $\pm$ 0.2\\
     60-70\%   &  $\geq$  29  & 31  $\pm$ 9  & 35   $\pm$ 13  & 7.1  $\pm$ 1.5 & 6.5  $\pm$ 1.6 & 0.39  $\pm$ 0.05  & 0.59  $\pm$ 0.07  & 2.3 $\pm$ 0.2\\
     70-80\%   &  $\geq$  13  & 16  $\pm$ 6  & 15   $\pm$ 8   & 4.8  $\pm$ 1.7 & 4.0  $\pm$ 1.8 & 0.38  $\pm$ 0.06  & 0.68  $\pm$ 0.09  & 1.9 $\pm$ 0.3\\
    \hline
    \end{tabular}
    }
    \end{center}
  \end{table*}

  Since the positions of nucleons fluctuate event-by-event, the principal axes 
  of the participant nucleons in the transverse plane are tilted and rotated with 
  respect to the original coordinate system. We define the participant plane (PP)
  which is the relevant plane to take into account the event-by-event position 
  fluctuations of participant nucleons.
  The transverse area and eccentricity with respect to the reaction plane (RP)
  and participant plane are defined as
  \begin{eqnarray}
    S_{\rm RP} & = & \pi \sqrt{\sigma_x^2 \sigma_y^2}, \\
    S_{\rm PP} & = & \pi \sqrt{\sigma_x^2 \sigma_y^2 - \sigma_{xy}^2}, \\
    \varepsilon_{\rm RP} & = & \frac{\sigma_y^2 - \sigma_x^2}{\sigma_y^2 + \sigma_x^2}, \\
    \varepsilon_{\rm PP} & = & \sqrt{ \frac{(\sigma_y^2 - \sigma_x^2)^2 + 4\sigma_{xy}^2}{\sigma_y^2 + \sigma_x^2} },
  \end{eqnarray}
  where 
  $\sigma_x^2 = \{x^2\} - \{x\}^2$,
  $\sigma_y^2 = \{y^2\} - \{y\}^2$
  and $\sigma_{xy} = \{xy\} - \{x\} \{y\}$.
  The curly brackets $\{...\}$ denote the average over all participants 
  for a given event.
  We have also calculated the averaged transverse path length $L$ from the RMS width
  \begin{equation}
    L = \sqrt{\sigma_x^2 + \sigma_y^2},
    \label{eq:pathlength}
  \end{equation}
  which could be a relevant geometrical quantity for the $R_{AA}$.
  The path length is very close to $\rho L$ defined in~\cite{Adler:2006bw},
  while Eq.~(\ref{eq:pathlength}) takes into account the event-by-event
  center of mass shift of the nuclei within the transverse plane.
  Average quantities, 
  $\langle\npart\rangle$, 
  $\langle\ncoll\rangle$,
  $\langle S\rangle$,
  $\langle\varepsilon\rangle$
  and $\langle L\rangle$
  have been calculated for each centrality bin
  where $\langle...\rangle$ describe the average over all events.

  \begin{figure}[htbp]
  \begin{center}
    \includegraphics[scale=0.7]{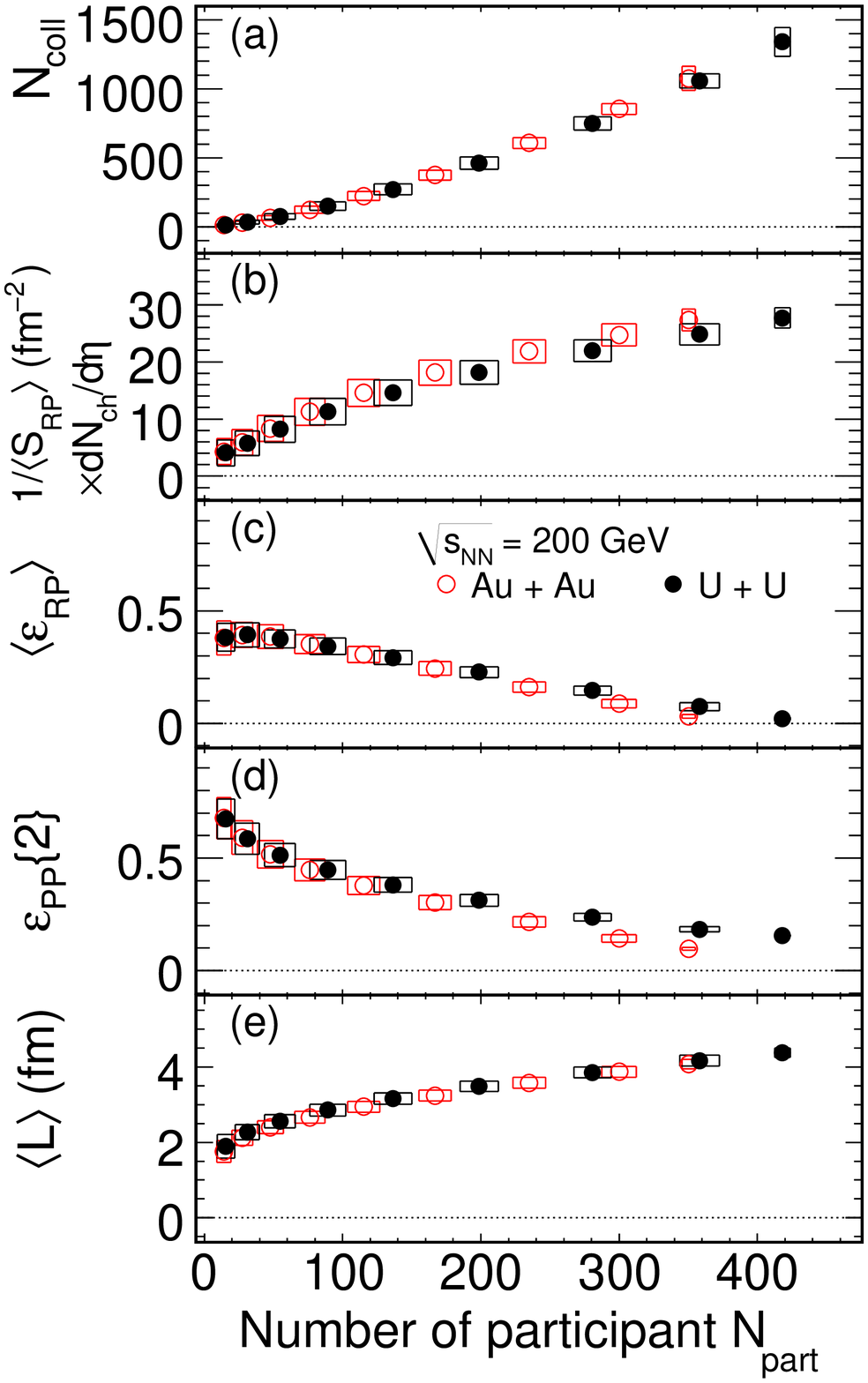}
    \caption{\label{fig:fig2}
      (Color online) Comparison of (a) \ncoll, (b) $1/\langle S_{\rm RP}\rangle \dNchdeta$,
      (c) $\langle\varepsilon_{\rm RP}\rangle$,
      (d) $\varepsilon_{\rm PP}\{2\}$
      and (e) $\langle L\rangle$
      as a function of \npart in \AuAu (open circles)
      and \UU collisions (filled circles) at \sqrts{200}.
      Open boxes show the systematic uncertainties in Monte Carlo Glauber simulations.
    }
  \end{center}
  \end{figure}
  Systematic uncertainties on the average quantities have been estimated 
  (i) by varying input parameters $R$, $a$, $n_{pp}$, $x$ as well as 
  the total cross section within $\pm$5\% 
  and (ii) by using different density profiles for nucleons
  in the Monte Carlo Glauber simulations. The dominant source of 
  systematic uncertainty is the total cross section.
  Total systematic uncertainty has been evaluated by the quadratic 
  sum of individual systematic uncertainty.
  Table~\ref{tab:summary_centrality} summarizes the centrality bins,
  average quantities and their systematic uncertainties obtained in the Monte Carlo Glauber simulation.

  Figure~\ref{fig:fig2} compares the \ncoll, 
  transverse number density $1/\langle S_{\rm RP}\rangle \dNchdeta$,
  reaction plane eccentricity $\langle\varepsilon_{\rm RP}\rangle$,
  second order cumulant of participant eccentricity $\varepsilon_{\rm PP}\{2\} \equiv \sqrt{\langle\varepsilon_{\rm PP}^2\rangle}$,
  and path length $\langle L\rangle$
  as a function of \npart together with their systematic uncertainties
  in \AuAu and \UU collisions at \sqrts{200}.
  We found that the all geometrical quantities essentially scale with \npart.
  The \ncoll, $\langle S\rangle$,  $\langle L\rangle$ increase 
  and $\langle\varepsilon_{\rm RP}\rangle$ decreases 
  at most central \UU collisions compared to those in \AuAu collisions
  because of the larger size of Uranium. 
  One can see that the $\varepsilon_{\rm PP}\{2\}$ in \UU collisions 
  starts deviating the \npart scaling around \npart = 200,
  and increases $\sim$60\% at top 5\% central \UU collisions. 
  The higher values of $\varepsilon_{\rm PP}\{2\}$ in \UU collisions 
  for large \npart is purely from the ground-state deformation of Uranium.
  We have confirmed that the $\varepsilon_{\rm PP}\{2\}$ becomes the same 
  if we assume the Uranium is spherical. 
  The relevance of the $\varepsilon_{\rm PP}\{2\}$ will be discussed in the next section.

\section{Results and Discussions}
\label{sec:results}


  \subsection{Elliptic flow $v_2$}
  \label{subsec:elliptic_flow}

    It has been found that the elliptic flow $v_2$ divided by initial anisotropy $\varepsilon$ in coordinate space
  scaled like $1/S \dNchdy$ among different energies and collision systems from AGS to RHIC~\cite{Voloshin:2007af}.
  A simple formula that has been proposed in~\cite{Bhalerao:2005mm} 
  describes very well the variation of $v_2$ with $1/S dN/dy$ 
  \begin{equation}
    \frac{v_2}{\varepsilon} = \frac{h}{1 + B~(1/S dN/dy)^{-1}},
    \label{eq:hydrolimit}
  \end{equation}
  where $dN/dy$ is the rapidity density of total particles,
  $h$ is the $v_2/\varepsilon$ in the ideal hydrodynamical limit when $1/S~dN/dy \rightarrow \infty$,
  and $B$ contains informations about the equation of state and the partonic cross section~\cite{Bhalerao:2005mm}. 
  The Eq.~(\ref{eq:hydrolimit}) reduces $v_2/\varepsilon \sim (h/B) 1/S~dN/dy$ when $1/S~dN/dy \rightarrow 0$
  for leading order in $1/S~dN/dy$, thus the above equation satisfies both low density and ideal 
  hydrodynamical limit of $v_2$.
  The integrated $v_2$ for unidentified charged hadrons from the PHOBOS collaboration 
  can be well described by the Eq.~(\ref{eq:hydrolimit})~\cite{Drescher:2007cd}.
  Assuming no change in the collision dynamics, we will study the $v_2/\epsilon$ distributions 
  versus the collision centrality in \UU collisions.

  \begin{figure}[htbp]
    \includegraphics[width=\linewidth]{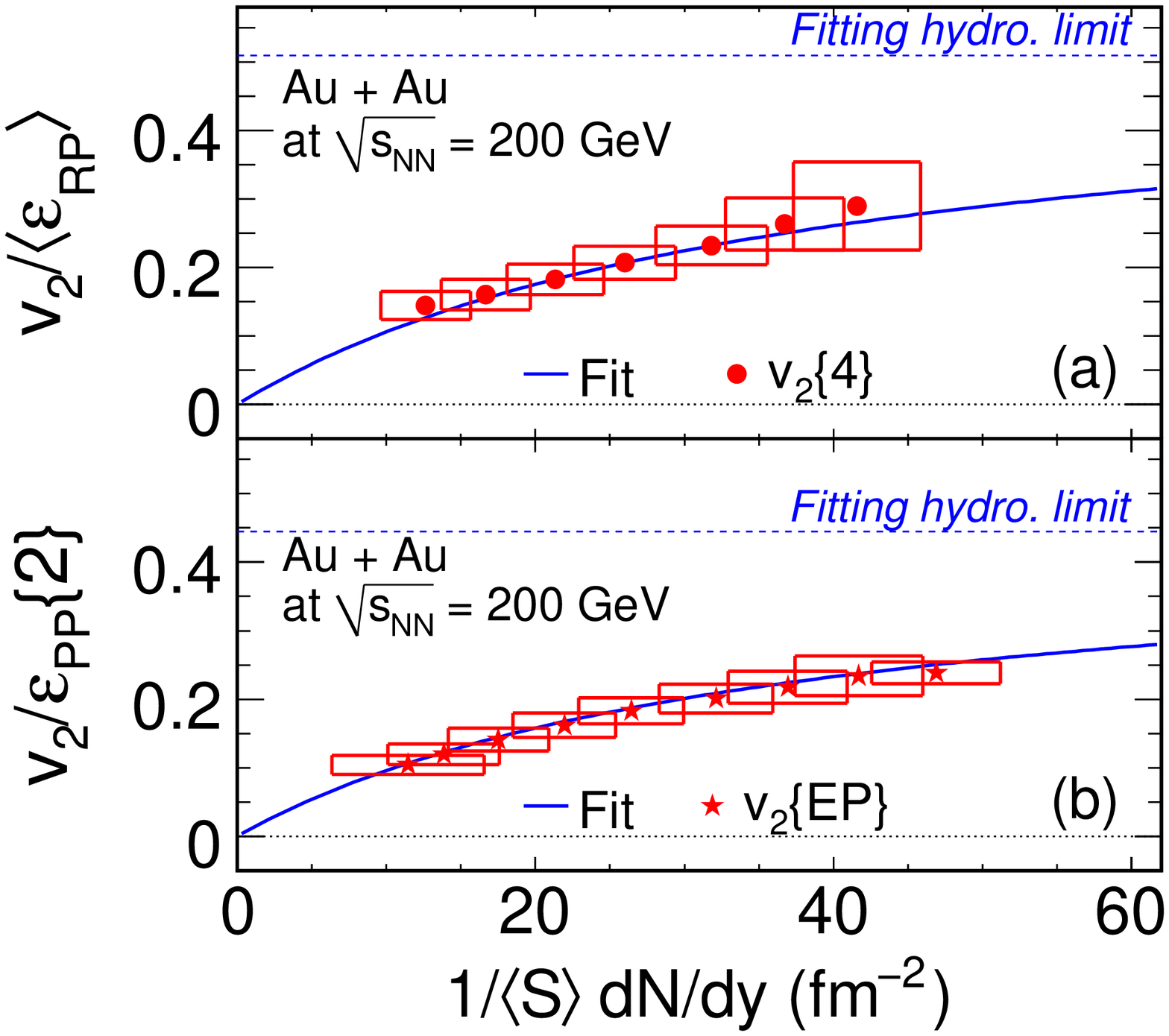}
    \caption{\label{fig:fig3}
      (Color online) (a) Four-particle cumulant $v_2\{4\}/\langle\varepsilon_{\rm RP}\rangle$ 
      as a function of $1/\langle S_{\rm RP}\rangle dN/dy$ for unidentified charged hadrons 
      in \AuAu collisions at \sqrts{200}.
      (b) The same plot as (a) for the standard event plane $v_2$\{EP\}$/\varepsilon_{\rm PP}\{2\}$ 
      as a function of $1/\langle S_{\rm PP}\rangle dN/dy$.
      Both the values of $v_2\{4\}$ and $v_2$\{EP\} are taken from~\cite{Adams:2004bi}. 
      Only statistical errors on the $v_2$ are shown and are smaller than symbols.
      Open boxes are systematic errors from the Monte Carlo Glauber simulation.
      Solid lines are fitting results by Eq.~(\ref{eq:hydrolimit}).
      See more details about fitting in the texts.
    }
  \end{figure}
    Figure~\ref{fig:fig3}
  shows the $v_2\{4\}/\langle\varepsilon_{\rm RP}\rangle$ and $v_2$\{EP\}$/\varepsilon_{\rm PP}\{2\}$ 
  as a function of $1/\langle S\rangle dN/dy$ in \AuAu collisions at \sqrts{200},
  where $v_2$\{4\} and $v_2$\{EP\} denote the $v_2$ from four particle cumulant method~\cite{Borghini:2000sa,Borghini:2001vi}
  and that from standard event plane method~\cite{Poskanzer:1998yz}, respectively.
  The $dN/dy$ is obtained by multiplying 3/2 to the measured \dNchdy at STAR~\cite{Abelev:2008ez}
  to take into account the neutral particles.
  The simultaneous fit has been performed for 
  (a) four particle cumulant $v_2\{4\}$, six particle cumulant and $q$-distribution method
  and for
  (b) standard event plane $v_2$\{EP\}, scalar product method, $\eta$ subevent, random subevent,
  and two particle cumulant. The results of $v_2$ are taken from~\cite{Adams:2004bi}.
  The two different groups of $v_2$ are categorized based on the multi-particle methods for (a)
  and two particle methods for (b). 
  As long as the distribution of eccentricity is 2D gaussian in the transverse plane, the effect of 
  fluctuation on the $v_2$\{4\} is negligible and thus the $\varepsilon_{\rm RP}$ can be used to 
  scale the $v_2$\{4\}~\cite{Voloshin:2007pc}.
  This assumption holds except for the peripheral 60-80\% centrality, where the distribution of 
  eccentricity becomes non-Gaussian.
  The $v_2$ from two particle methods are expressed as $(v_2^{\alpha})^{1/\alpha}$ where 
  $v_2$ is the true $v_2$ value and $\alpha$ varies from 1 to 2 depending on the event 
  plane resolution~\cite{Alver:2008zza,Ollitrault:2009ie}.
  In this study, $\varepsilon_{\rm part}\{2\} = \sqrt{\langle \varepsilon_{\rm part}^2\rangle}$
  was used by assuming $\alpha = 2$. We confirmed that the resulting $v_2$ values unchanged 
  by using $\varepsilon_{\rm part}$ (i.e. $\alpha = 1$). Because the $v_2$ values were 
  extrapolated from $(v_2/\varepsilon)$ multiplied by $\varepsilon$, most of the difference 
  between $\varepsilon_{\rm part}\{2\}$ and $\varepsilon_{\rm part}$ is canceled out and thus 
  the resulting $v_2$ is the same regardless of the choice of eccentricity.

  \begin{figure}[htbp]
    \includegraphics[width=\linewidth]{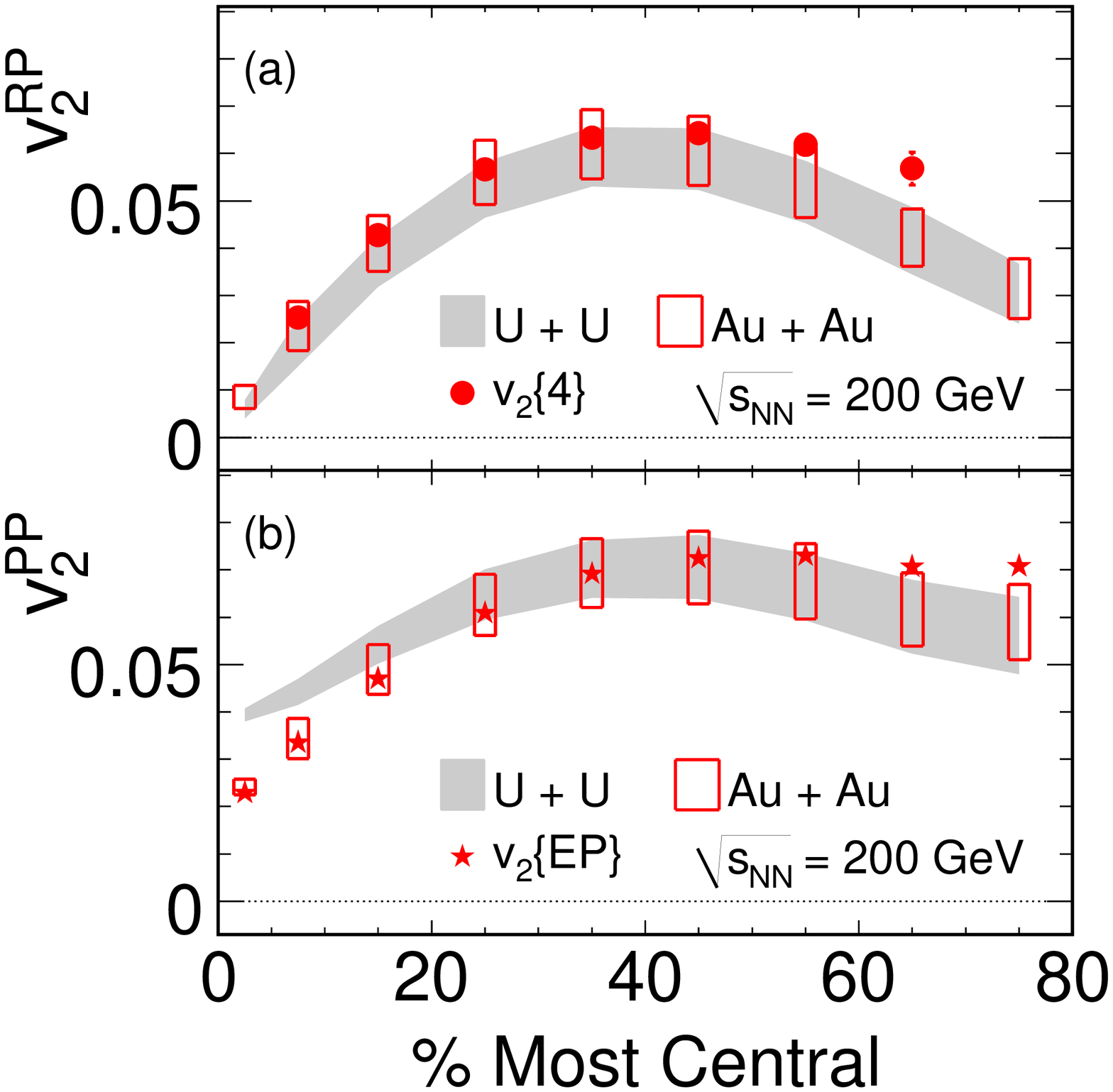}
    \caption{\label{fig:fig4}
      (Color online) (a) The $v_2^{\rm RP}$ as a function of centrality.
      Solid circles is the $v_2\{4\}$, open boxes and shaded band show 
      the extracted $v_2$ from the fitting of $v_2/\langle\varepsilon_{\rm RP}\rangle$ and 
      $\langle\varepsilon_{\rm RP}\rangle$ in Monte Carlo Glauber simulation
      in \AuAu and \UU collisions, respectively.
      (b) The same comparison of $v_2^{\rm PP}$ with $\varepsilon_{\rm PP}\{2\}$ as (a), 
      where solid stars are the $v_2$\{EP\}.
      The errors on the $v_2$ include systematic errors from Monte Carlo Glauber simulation
      and errors from fitting of $v_2/\varepsilon$.
    }
  \end{figure}
  Figure~\ref{fig:fig4} shows the extracted $v_2$
  in \UU collisions compared to those in \AuAu collisions 
  as a function of centrality at \sqrts{200}.
  The $v_2^{\rm RP}$ ($v_2^{\rm PP}$) denotes the $v_2$ measured 
  with respect to the reaction (participant) plane.
  The $v_2^{\rm RP}$ and $v_2^{\rm PP}$ have been calculated by multiplying 
  the $\langle\varepsilon_{\rm RP}\rangle$ and $\varepsilon_{\rm PP}\{2\}$
  to the fitting results of $v_2/\varepsilon$ shown in Fig.~\ref{fig:fig3}
  for each centrality bin.
  Since we have calculated the \dNchdeta in the Monte Carlo Glauber simulation,
  it is necessary to convert the \dNchdeta to $dN/dy$ for calculating the $v_2$
  for each centrality bin. 
  We assume that $\dNchdy \approx 1.15 \dNchdeta$
  to extrapolate the $v_2/\varepsilon$ for each centrality~\cite{Adler:2002pu}.
  The $v_2^{\rm RP}$ in \UU collisions is consistent with that in \AuAu 
  collisions for centrality 0-80\%. The $v_2^{\rm PP}$ is also consistent 
  with each other in \UU and \AuAu collisions for centrality 20-80\%,
  whereas the $v_2$ in \UU collisions at top 0-10\% centrality is 30-60\% 
  larger than that in \AuAu collisions. The larger $v_2$ is attributed to
  the larger participant eccentricity due to the ground-state deformation 
  in top 0-10\% centrality in \UU collisions compared to that in \AuAu collisions.
  The extracted $v_2$ in \AuAu collisions are slightly smaller than the 
  data at peripheral collisions. Since the \dNchdeta has been tuned
  to reproduce the PHOBOS results and is smaller than the STAR \dNchdy
  at peripheral 60-80\%, the resulting $v_2/\varepsilon$ (and hence the $v_2$) 
  become smaller than the STAR $v_2$.

  \subsection{Nuclear modification factor $R_{AA}$}
  \label{subsec:raa}

  The integrated $R_{AA}$ over a certain \pt range 
  in \AuAu collisions at \sqrts{200} has been described by
  $R_{AA}=(1-S_0\npart^a)^{n-2}$,
  where $n = 8.1$ is the power-law exponent of \pt distribution,
  and $S_0 = (9.0 \pm 6.1) \times 10^{-3}$ and
  $a = 0.57 \pm 0.14$ for $\npart > 20$ and \pt $> 5$ \gevc~\cite{Adare:2008qa}.
  We have assumed that the path length $\langle L\rangle$ determines the $R_{AA}$ 
  in both \AuAu and \UU collisions. 
  The $R_{AA}$ in \UU collisions has been extrapolated by fitting the $R_{AA}(L)$ in Au + Au collisions
  with an ansatz from above equation
  \begin{equation}
    R_{AA}(L) = (1 - S_0' \langle L\rangle^{b})^{n-2},
    \label{eq:raa_L}
  \end{equation}
  where $n = 8.1$, $S_0'$ and $b$ are free parameters that have been evaluated by 
  fitting the data.

  \begin{figure}[htbp]
    \includegraphics[width=\linewidth]{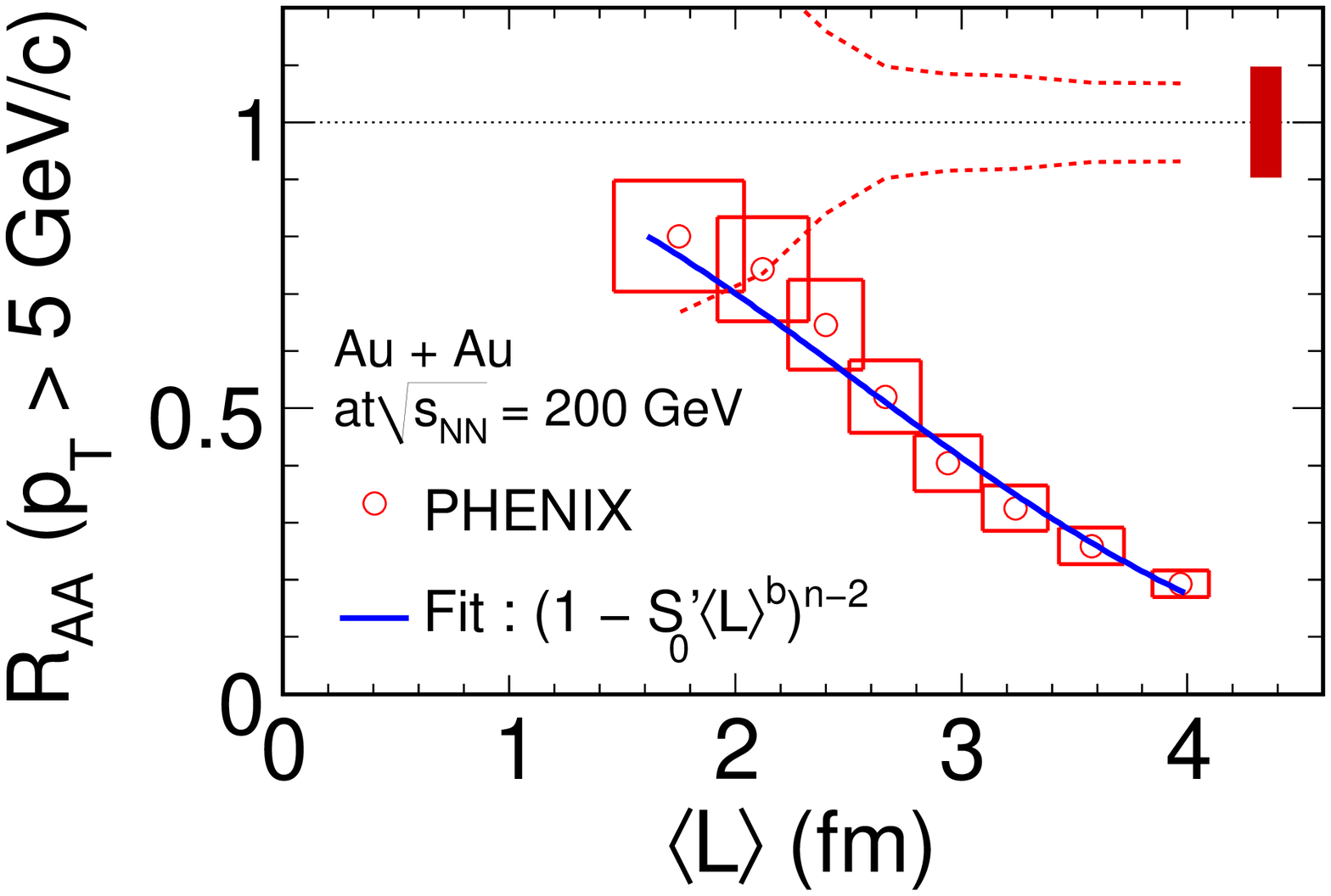}
    \caption{\label{fig:fig5}
      (Color online) Integrated $R_{AA}$ above \pt~$>$~5~\gevc~\cite{Adare:2008qa}
      as a function of $\langle L\rangle$ in \AuAu collisions at \sqrts{200}.
      Most central bin denote 0-10\% rather than 0-5\%.
      Statistical and \pt-uncorrelated errors are smaller than symbols.
      Open boxes are the errors on $\langle L\rangle$ in x axis and 
      \pt-correlated systematic errors on the $R_{AA}$ in y axis.
      The dashed lines and the single box on right at unity show the 
      errors on \ncoll and normalization of the \pp reference, respectively~\cite{Adare:2008qa}.
      Solid line is the fitting result by Eq.~(\ref{eq:raa_L}).
    }
  \end{figure}
    Figure~\ref{fig:fig5} shows integrated $R_{AA}$ for \pt~$>$~5~\gevc from~\cite{Adare:2008qa} 
  as a function of $\langle L\rangle$ in \AuAu collisions at \sqrts{200}. 
  We have assumed that the definition of our centrality bins is 
  the same as that of the PHENIX in order to plot the $R_{AA}$ as a function 
  of $\langle L\rangle$ for each centrality. 
  Result of the fit with Eq.~(\ref{eq:raa_L}) is shown by the solid line 
  in Fig.~\ref{fig:fig5} and holds quite well over entire range of $\langle L\rangle$
  since the $\langle L\rangle$ scales like $\npart^{1/3}$.

  \begin{figure}[htbp]
    \includegraphics[width=\linewidth]{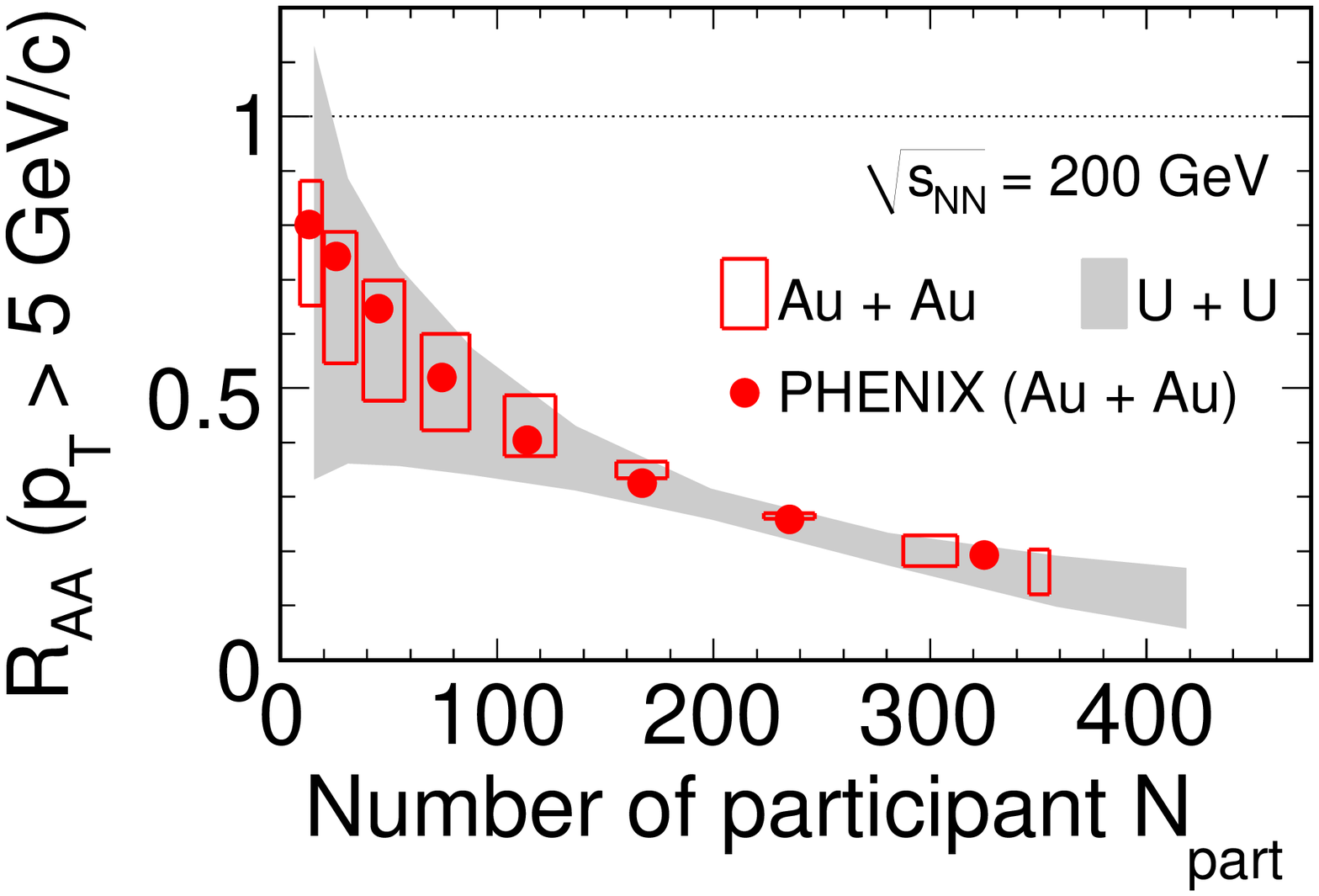}
    \caption{\label{fig:fig6}
      (Color online) Integrated $R_{AA}$ as a function of \npart at \sqrts{200},
      where open boxes and shaded bands show the $R_{AA}$ extrapolated by Monte Carlo Glauber simulation
      in \AuAu and \UU collisions, respectively.
      Fitting error and systematic errors from Monte Carlo Glauber simulation have been included.
      Additional error from \ncoll has also been included in \UU collisions.
      The result in \AuAu at \sqrts{200} from PHENIX experiment (solid circles)~\cite{Adare:2008qa}
      are plotted for comparison.
    }
  \end{figure}
    Figure~\ref{fig:fig6} shows integrated $R_{AA}$ as a function of \npart 
  extrapolated for \UU collisions at \sqrts{200}.
  The $R_{AA}$ in \UU collisions has been evaluated 
  for a given $\langle L\rangle$ in each centrality.
  The calculated $R_{AA}$ in \AuAu collisions (open boxes) is consistent 
  with the data within the systematic error as it should.
  We found that the $R_{AA}$ reaches $\sim$0.1 at most central \UU collisions,
  which is by a factor of 2 more suppression compared to the central \AuAu collisions
  due to the larger size and the deformation of Uranium.
  Heinz and Kuhlman pointed out in~\cite{Heinz:2004ir} that 
  the radiative energy loss $\Delta E$ of a fast parton moving through the medium
  is almost independent of the orientations of nuclei for the out-of-plane direction
  in the full overlap U~+~U collisions.
  Whereas the $\Delta E$ in the in-plane direction decreases by about 
  35\% towards the ideal body-body collisions (head-on collisions along the shortest axes).
  For the body-body collisions, they found that the difference of $\Delta E$ 
  between out-of-plane and in-plane directions is more than twice in U~+~U collisions 
  that achieved in Au~+~Au collisions.
  They also found that the total energy loss is larger by up to a factor of 2.
  More differential study, such as selecting the orientations of Uranium
  and directions with respect to the reaction plane, will be needed to 
  see whether the $R_{AA}$ would have such dependences or not.
  Since the $\langle L\rangle$ in \UU collisions is slightly larger ($\sim$3\% in central, 
  and $\sim$5\% in peripheral collisions) than that in \AuAu collisions,
  the $R_{AA}$ in \UU  collisions would be even more suppressed for a given \npart.
  Due to the large errors on the extrapolated $R_{AA}$, we have not observed any 
  difference of the $R_{AA}$ between \AuAu and \UU collisions for a given \npart.

\section{Summary}
\label{sec:summary}

  In summary, we have predicted the $v_2$ and $R_{AA}$
  in \UU collisions at \sqrts{200} by a simple geometrical 
  approach with the Monte Carlo Glauber simulation.
  We found that the $v_2^{\rm RP}$ is consistent with that 
  in \AuAu collisions over all centrality range, whereas 
  the $v_2^{\rm PP}$ increase by 30-60\% at most central 0-10\% collisions
  due to the larger $\varepsilon_{\rm PP}\{2\}$ in \UU collisions.
  The $R_{AA}$ at top 5\% central \UU collisions further 
  suppressed and reaches $\sim$0.1, which is by a factor of 2 more 
  suppression compared to the most central \AuAu collisions.
  It is clear that the larger mass and deformation form the U nucleus 
  will allow us to study the matter at higher density. 
  By selecting the relative orientation of the colliding Uranium nuclei, 
  the discussed effects may be further enhanced. 
  We will report the method in a separate paper.

\section*{Acknowledgements}

  We thank Peter Filip and Art Poskanzer for discussions.
  The work is supported in part by the U.S. Department of Energy under Contract No. DE-AC03-76SF00098.




\end{document}